\documentclass[twocolumn,prb]{revtex4}
\usepackage{amsmath,graphicx}

\begin{document}
\title{Rotating Superconductors and the London Moment:\\
Thermodynamics versus Microscopics}
\author{Yimin Jiang$^{(a,b)}$ and Mario
Liu$^{(a)}$~\cite{email}}
%\address
\affiliation {(a) Institut f\"{u}r Theoretische Physik,
Universit\"{a}t Hannover,\\ 30167 Hannover, Germany \\ (b)
Department of Applied Physics and Heat-Engineering,\\
Central South University of Technology, Changsha 410083,
P. R. China}
\date{\today}

\begin{abstract}
Comparing various microscopic theories of rotating
superconductors to the conclusions of thermodynamic
considerations, we traced their marked difference to the
question of how some thermodynamic quantities (the
electrostatic and chemical potentials) are related to
more microscopic ones: The electron's the work function,
mean-field potential and Fermi energy -- certainly a
question of general import.

After the correct identification is established, the
relativistic correction for the London Moment is shown to
vanish, with the obvious contribution from the Fermi
velocity being compensated by other contributions such as
electrostatics and interactions.
\end{abstract}

\maketitle
\section{Introduction}
The defining property of superconductors, the well known
Mei{\ss}ner Effect, is to expel an (undercritical) magnetic
field from its bulk. This property reverses itself when
the superconductor is rotated, and a spontaneous magnetic
field appears -- again in the bulk, one or two
penetration lengths away from the surface. This is
usually referred to as the London Moment~\cite {london}.
Its magnitude is~\cite{HL}
\begin{equation}  \label{0}
{\bf B}= -[2m_e(1+\zeta)c/e]{\boldsymbol\Omega},
\end{equation}
where $\Omega$ is the rotational velocity, $c$ the vacuum
light velocity, $m_{\mathrm{e}}$ the bare mass, and $e<0$
the charge, of the electron; finally, $\zeta$ is a
relativistically small correction. All theories and
experiments agree that $\zeta$ is small, but strong
disagreement exists with respect to its actual value and
sign. The microscopic theories, by Anderson~\cite{pwa},
Brady~\cite{brady}, Cabrera and
coworkers~\cite{gutfr,peskin}, and Baym~\cite{baym} take
the main correction to be positive and given by the Fermi
velocity,
\begin{equation}  \label{m}
\zeta\approx (v_{\mathrm{F}}/c)^2\approx2\times10^{-4}.
\end{equation}
The thermodynamic theory, on the other hand, finds $\zeta$ to be negative,
and very much smaller~\cite{liu}:
\begin{equation}  \label{M}
\zeta= \tilde\mu/c^2\approx-10^{-10},
\end{equation}
where $\tilde\mu$  is the chemical potential of the
metal, the energy needed to add a unit mass to the
superconducting solid. (Since the solid holds its atoms
together, the chemical potential $\tilde\mu$ is
necessarily a negative quantity.) The value of $10^{-10}$
makes the correction negligible for all conceivable
purposes and renders Eq(\ref{0}), taking the values of
$e$ and $c$ as given, a very precise expression for $m_e$
-- about three orders of magnitude more precise than any
present experimental technique to determine the electron
mass directly~\cite{d}.

Generally speaking, although the thermodynamic,
macroscopic theory lacks details and is incapable of
answering many quantitative questions, it is nevertheless
a rigorous theory, and uniquely appropriate for
understanding the London moment. This is because (i) the
London Moment is an equilibrium phenomenon, (ii) the
measurement concerns a quotient between two macroscopic
fields, $B$ and $\Omega$, and most importantly, (iii) the
reason for the microscopic parameters of $m_{\mathrm{e}}$
and $e$ to appear on the macroscopic level is due to
simple symmetry principles, rather than some detailed
properties of the interaction.

Comparing both types of theories in detail, we find their
difference to be easy to bridge, if the relations between
some thermodynamic and microscopic quantities had been
clear beforehand. More specifically, it is the incorrect
identification of electrostatic and chemical potentials
on one hand, the work function, mean-field potential and
Fermi energy of the electron on the other that has led to
the above discrepancy. The correct identification of
these quantities is of rather general interest and
transcends the understanding of the London Moment alone.
The proper identification is the most important result of
the present paper.

Another problem clouding the understanding of the London
Moment is the disagreement between experiment and all
theories. Although this remains a point we do not
understand, we took a step towards its clarification by
achieving understanding of the following point: The best
present experiment measures the flux, $\oint
\mathbf{A}\cdot \mathrm{d}\mathbf{s}$ $=\int
\mathbf{B}\cdot \mathrm{d}\mathbf{a}$, outside the
rotating superconductor~\cite{cabr}. Two effects
contribute the this flux, the London Moment, and the
so-called double layer. The latter is a result of the
fact that a metal may be conceived of as an electrostatic
potential of the square-well form, with the depth $\Delta
\Phi $. As the metal rotates, this discontinuity in the
rest-frame electrostatic potential produces a
discontinuity in the laboratory-frame vector potential,
\begin{equation}
\Delta {\mathbf{A}}=-({\boldsymbol\Omega}
\times{\boldsymbol r})\Delta \Phi /c.  \label{dl}
\end{equation}
And this also contributes to the flux $\oint
\mathbf{A}\cdot \mathrm{d}  \mathbf{s}$, with a
contribution much larger than the relativistic correction
of the London Moment. It has been generally suspected
that both contributions are experimentally inseparable,
that they are always measured in conjunction. (A double
layer of opposite charge -- or one layer of dipoles --
produces a discontinuity in the electrostatic potential,
hence the name.)

We disagree. In this paper, we shall discuss both effects
separately, as only the first is universal -- the second
varies with materials, and is quite independent from
superconductivity. Besides, we believe that these effects
may in principle be measured independently. Taking the
superconductor to be a cylinder rotating around its axis,
the London Moment may be measured immediately outside its
top or bottom surface, where any magnetic field
$\mathbf{B}$ is continuous, see Fig~1. (It is not
continuous across the cylinder surface, where the field
drops to zero within a few penetration lengths, due to
the presence of persistent currents.) The flux from the
double layer alone  may be seen above
$\mathrm{T_\lambda}$, or in the experiment proposed in
chapter~\ref{exp}.
\begin{figure}
\begin{center}
\includegraphics[width=0.6\columnwidth]{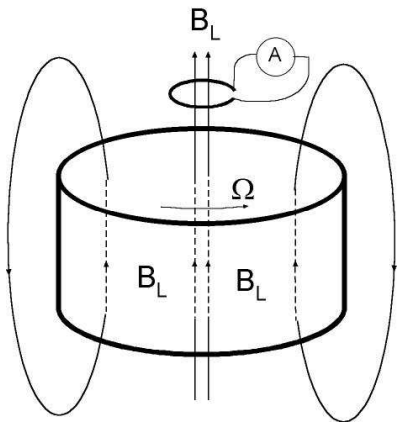}
\end{center}
\caption{Direct measurement of the London field
$B_L$.}\label{Fig1}
\end{figure}
In chapter~\ref{1}, we shall present the simple yet
stringent macroscopic calculation leading to the London
Moment. This is followed in chapter~\ref{3} by a
comparison to the microscopic results, which are brought
into agreement with the macroscopic ones by the
appropriate identification of the above mentioned
thermodynamic and microscopic quantities. In chapter
~\ref{4}, we consider the effect of the double layer, the
contribution of which may be measured in the experiment
described in chapter~\ref{exp}.

\section{The Macroscopic Approach}\label{1}
\subsection{Electrostatic and Chemical Potential}
\label{1.1} Given any neutral, macroscopic system, we may
either change its numbers of electrons and ions, $N_{-}$
and $N_{+}$, or its mass and charge, $M$ and $Q$. They
are related as
\begin{eqnarray}
M &=&m_{\mathrm{i}}N_{+}+m_{\mathrm{e}}N_{-},  \label{a1} \\
Q &=&|e|(N_{+}-N_{-}).  \label{1a}
\end{eqnarray}
($m_{\mathrm{i}}$ and $m_{\mathrm{e}}$ denote the bare
mass of the ion and electron, respectively, and $e<0$ is
the elementary charge.) Frequently, the number of atoms
$N_{\mathrm{a}}\equiv M/(m_{\mathrm{i}}+m_{\mathrm{e}})$
is used instead of the mass $M$. The energy of these
changes are given by the respective chemical potentials,
\begin{eqnarray}
\mathrm{d}E &=&\mu _{+}\mathrm{d}N_{+}+\mu _{-}\mathrm{d}
N_{-}  \label{a2} \\ &=&\mu \mathrm{d}M+\Delta \Phi
\mathrm{d}Q  \label{a3} \\ &=&\mu
_{\mathrm{a}}\mathrm{d}N_{\mathrm{a}}+\Delta \Phi
\mathrm{d}Q, \label{a4}
\end{eqnarray}
where employing Eqs~(\ref{a1},\ref{1a}), we see the
chemical potentials to be related as
\begin{eqnarray}
\mu _{+} &=&m_{\mathrm{i}}\mu -e\Delta \Phi ,  \label{5}
\\ \mu _{-} &=&m_{\mathrm{e}}\mu +e\Delta \Phi ,
\label{6} \\ \mu &=&\mu
_{\mathrm{a}}/(m_{\mathrm{i}}+m_{\mathrm{e}}). \label{6a}
\end{eqnarray}
Because $\mu _{-}$, $\mu _{+}$, $\mu _{\mathrm{a}}$, $\mu
$ and $\Delta \Phi $ respectively denote the energy
needed to bring an electron, an ion, an atom, unit mass
and unit charge from infinity to inside the system, these
are fixed quantities, and we are not at liberty to alter
them by an additive constant. We especially have
\begin{equation}
\Delta \Phi =\Phi ({\boldsymbol r})-\Phi (\infty ),  \label{7}
\end{equation}
where $\Phi $ is the electrostatic potential, which is
fixed only up to an arbitrary gauge transformation, and
which forms a 4-vector with the vector potential
$\mathbf{A}$. Note that the energy and chemical
potentials employed in this paper include the rest mass,
because otherwise the relativistic transformation
formulas of the following sections will not work. Hence
we define the non-relativistic chemical potentials with
tilde,
\begin{eqnarray}
\mu _{+} &=&m_{\mathrm{i}}c^2+\tilde{\mu}_{+},  \label{7c} \\
\mu _{-} &=&m_{\mathrm{e}}c^2+\tilde{\mu}_{-},  \label{7d} \\
\mu &=&c^2+\tilde{\mu},  \label{7e} \\
\mu _a &=&(m_{\mathrm{i}}+m_{\mathrm{e}})c^2+\tilde{\mu}_a  \label{7f}
\end{eqnarray}
Usually, both $\tilde\mu_{\mathrm{a}}$ and $\tilde\mu_-$
are of the order of a few eV, this being the scale of
atomic physics, cf an early model-calculation~\cite{wig}.
Since $m_{\mathrm{e}}/(m_{\mathrm{i}}+m_{\mathrm{e}})
\approx10^{-5}$, Eq~(\ref{6}) and (\ref{6a}) imply, to
great accuracy,
\begin{equation}  \label{7b}
\tilde\mu_-=e\Delta\Phi,
\end{equation}
with
\begin{equation}  \label{7a}
\tilde\mu_-/m_{\mathrm{e}}\approx10^{-5}c^2, \qquad
\tilde\mu\approx10^{-10}c^2.
\end{equation}

\subsection{Josephson Equation and Superfluid Velocity}
\label{1.2}

The Josephson equation is usually given as
\begin{equation}  \label{8}
(\hbar/2)\dot\varphi+\mu_-=0,
\end{equation}
though this equation is not gauge invariant. In view of
Eq~(\ref{6}) and (\ref {7}), the correct form for the
Josephson equation is clearly
\begin{equation}  \label{9}
(\hbar/2)\dot\varphi+m_{\mathrm{e}}\mu+e\Phi=0,
\end{equation}
implying that Eq~(\ref{8}) is valid only if
$\Phi(\infty)=0$, a special gauge choice. (Similarly,
taking $\dot\varphi=0$ when considering Eq~(\ref{9}) also
constitutes a gauge choice.) Note that the form of
Eq~(\ref{9}) does presume rest frame of the crystal, and
equilibrium. The superfluid velocity is also defined as a
gauge invariant quantity
\begin{equation}  \label{10}
{\boldsymbol v^s}\equiv\frac{1}{m_{\mathrm{e}}} (
\frac{\hbar}{2}{\boldsymbol\nabla}\varphi-
\frac{e}{c}\mathbf{A}).
\end{equation}
Its equation of motion reads
\begin{equation}  \label{11}
\dot{\boldsymbol v^s}+
{\boldsymbol\nabla}\mu=e{\mathbf{E}}/m_{\mathrm{e}},
\end{equation}
obtained by inserting Eq~(\ref{10}) in (\ref{9}) and
employing ${\mathbf{E}}=-(\boldsymbol{\nabla}
\Phi+\dot{\mathbf{A}}/c)$. Although this pleasingly
simple equation, independent from any coefficients, may
be used to justify the definitions of
Eqs~(\ref{9},\ref{10}), we must be aware that
${\boldsymbol v^s}$ as defined need not transform as a
velocity under a Galilean-Lorentz boost. The
transformation behavior of ${\boldsymbol v^s}$ is given
by the Josephson equation (\ref{9}): Define a 4-velocity
\begin{eqnarray}
u_\alpha &\equiv &\frac \hbar
{2m_{\mathrm{e}}}\frac{\partial \varphi }{\partial
x^\alpha }-\frac e{m_{\mathrm{e}}c}A_\alpha  \label{13}
\\ &=&\frac \hbar {2m_{\mathrm{e}}}\left(
\dot{\varphi}/c,{\boldsymbol\nabla \varphi }\right)
-\frac e{m_{\mathrm{e}}c}\left( -\Phi ,\mathbf{A}\right)
,
\end{eqnarray}
and use Eq~(\ref{9}) to yield
\begin{equation}
u_\alpha =(-\mu /c,\,{\boldsymbol v^s}).  \label{14}
\end{equation}
Now, because $u_\alpha $ transforms as
\begin{equation}
u_\alpha ^{\prime }=(u_0-{\mathbf{u}}\cdot {\boldsymbol
v}/c,\,{\mathbf{u}}-u_0{\boldsymbol v}/c)  \label{15}
\end{equation}
under a boost of $v$,  to linear order in $v/c$, so does
the 4-vector $(-\mu /c,{\boldsymbol v^s})$, leading to
\begin{eqnarray}
\mu ^{\prime } &=&\mu + {\boldsymbol v^s}\cdot
{\boldsymbol v},  \label{16} \\ ({\boldsymbol
v^s})^{\prime } &=&{\boldsymbol v^s}+(\mu
/c^2){\boldsymbol v}, \label{17}
\end{eqnarray}
implying that it is the quantity
\begin{equation}
{\boldsymbol v^{ts}}\equiv  (\mu /c^2)^{-1}{\boldsymbol
v^s}={\boldsymbol v^s} /(1+\zeta )  \label{17a}
\end{equation}
that transforms as a velocity, $({\boldsymbol
v^{ts}})^{\prime }=$ ${\boldsymbol v^{ts}}+{\boldsymbol
v}$. We shall refer to ${\boldsymbol v^{ts}} $ as the
\textbf{t}rue \textbf{s}uperfluid velocity -- although
with
\begin{equation}
\zeta =\tilde{\mu}/c^2=(\tilde{\mu}_{-}-e\Delta \Phi
)/m_{\mathrm{e} }c^2\approx 10^{-10},  \label{MM}
\end{equation}
cf Eqs~(\ref{6},\ref{7e},\ref{7a}), the difference is
very small indeed. Defining an effective mass
$m_{\mathrm{e}}^{*}$ by
\begin{eqnarray}
m_{\mathrm{e}}^{*} &\equiv &m_{\mathrm{e}}(1+\zeta ),
\label{19} \\ {\boldsymbol v^{ts}}&=&\frac
1{m_{\mathrm{e}}^{*}}(\frac \hbar 2{\boldsymbol \nabla
}\varphi -\frac ec\mathbf{A}),  \label{19a}
\end{eqnarray}
we shall find in the next section that this is  the mass
appearing in the formula for the London Moment,
Eq~(\ref{0}).

\subsection{Thermo- and Hydrodynamics}
\label{1.3} In a general inertial frame, the thermodynamics of a
superconducting solid is given as
\begin{eqnarray}
\mathrm{d}\varepsilon &=&T\,\mathrm{d}s+\mu \,\mathrm{d}\varrho +{%
\boldsymbol v}^n\cdot \mathrm{d}{\boldsymbol g}+ \sigma
_{ij}\mathrm{d}\nabla _ju_i  \notag \\ &&+\mathbf{E}\cdot
\mathrm{d}\mathbf{D}+\mathbf{H} \cdot
\mathrm{d}\mathbf{B}+ \mathbf{j}^s\cdot
\mathrm{d}{\boldsymbol v}^s.  \label{20}
\end{eqnarray}
Notation and explanation: $s$ is the entropy density, and
$T$ the temperature. $\mu \,\mathrm{d}\varrho $ and
$\mathbf{E}^0\cdot \mathrm{d} \mathbf{D}$ are the
respective local expressions for $\mu \mathrm{d}M$ and
$\Delta \Phi \mathrm{d}Q$ of Eq~(\ref{a3}), where
$\mathbf{E}\cdot \mathrm{d} \mathbf{D}$ involves a
partial integration, and is manifestly gauge invariant;
$\mathbf{H}\cdot \mathrm{d}\mathbf{B}$ is the magnetic
counter term. $\mathbf{j}^s\cdot \mathrm{d}{\boldsymbol
v}^s$ characterizes the broken phase invariance of
superconductivity, as $\sigma _{ij}\mathrm{d}\nabla
_ju_i$ the broken translational invariance of solids --
with $u_i$ the displacement vector. In a general inertial
frame, the total, conserved momentum density
${\boldsymbol g}^{\mathrm{tot}}$ is also a thermodynamic
variable, though we employ ${\boldsymbol g}\equiv $
${\boldsymbol g}^{\mathrm{tot}}$ $-\mathbf{D\times B}$
instead~\cite{katja}, with ${\boldsymbol v^n}\equiv
\partial \varepsilon /\partial {\boldsymbol g }$ being
the equilibrium velocity of crystal points, atoms and
normal electrons. In the present case of interest,
\begin{equation}
{\boldsymbol v^n}={\boldsymbol\Omega }\times {\boldsymbol
r}.  \label{omega}
\end{equation}
Being a conjugate variable, $\mathbf{j}^s\equiv\partial
\varepsilon /\partial {\boldsymbol v^s}$ is a function of
those two thermodynamic variables also odd under time
reversal, $\boldsymbol v^s$ and $\boldsymbol v^n$. In an
expansion, to linear order of the variables, we have
\begin{equation}
j_i^s=(c^2/\mu )\rho _{ij}^s(v_j^s-\alpha _{jk}v_k^n),  \label{24}
\end{equation}
where $\rho _{ij}^s$ and $\alpha _{jk}$ are two expansion
coefficients, while the prefactor $(c^2/\mu )$ simply
renormalizes $\rho _{ij}^s$. A Maxwell relation then
implies
\begin{equation}
\left( \frac{\partial g_i}{\partial j_j^s}\right)
_{v_n}=\left( \frac{\partial v_j^s}{\partial
v_i^n}\right) _{j_s}=\alpha _{ji}.  \label{25}
\end{equation}
Confining ourselves  to the local rest frame and
disregarding dissipative terms (then
$\dot{s},\,\dot{u}_i=0$), the hydrodynamic set of
equations is given by the Josephson equation~(\ref{11}),
the Maxwell equations, and the conservation laws for
energy and mass
\begin{eqnarray}  \label{max1}
\dot{\varepsilon} &=&-{\boldsymbol\nabla} \cdot
{\mathbf{Q}},\qquad \dot{\varrho}+{\boldsymbol\nabla}
\cdot {\mathbf{j}}_\rho =0, \label{conti} \\
\dot{\mathbf{B}} &=&-c{\boldsymbol\nabla} \times {\mathbf
E},\quad \dot{\mathbf{D}} =c{\boldsymbol\nabla} \times
{\mathbf H}-{\mathbf j}_e,
\end{eqnarray}
where $\mathbf{Q}$, $\mathbf{j}_\rho $ and
$\mathbf{j}_{\mathrm{e}}$ are as yet unknown. Inserting
these expressions into the temporal derivative of
Eq~(\ref{20}), $\dot{\varepsilon}=$ $\mu \dot{\varrho}$
$+\mathbf{E}\cdot \dot{\mathbf{D}}$ $+\mathbf{H}\cdot
\dot{\mathbf{B}}+$ $\mathbf{j}^s\cdot \dot{\boldsymbol
v}^s$, and insisting that all equations are satisfied
simultaneously, we find
\begin{eqnarray}
{\mathbf j}_\rho &=&{\mathbf j}_s,\quad {\mathbf
j}_{\mathrm{e}}=e{\mathbf j}_s/m_{\mathrm{e}},
\label{erho} \\ {\mathbf Q} &=&{ \mathbf j}_s\mu
+c{\mathbf E}\times {\mathbf H}.  \label{QQ}
\end{eqnarray}
Clearly, $\mathbf{j}_s\equiv $ $\partial \varepsilon
/\partial {\boldsymbol v^s}$ has the significance of
being the persistent mass and electric current in the
rest frame. The relation
\begin{equation}
{\mathbf{j}}_\rho =e{\mathbf{j}}_s/m_e,  \label{22}
\end{equation}
is a necessarily one, as the transfer of one electron is
coupled to the transfer of the bare values of $e$ and
$m_{\mathrm{e}}$. Starting from this relation and tracing
it back, we would have found that the charge $e$ in
Eq~(\ref{10}) must indeed be the bare charge, and that
${\boldsymbol v^s}$ must indeed be gauge invariant.
Because of the symmetry of the energy stress 4-tensor,
${\boldsymbol g}^{\mathrm{tot}}$ $={\mathbf Q}/c^2$, we
have
\begin{equation}
{\boldsymbol g}^{\mathrm{tot}}={\boldsymbol g}={\mathbf
j}_s\mu /c^2, \label{23}
\end{equation}
for $E$, $D=0$ (and still in the local rest frame). This implies
\begin{equation}
\left( \frac{\partial g_i}{\partial j_j^s}\right) _{v_n}=\frac \mu
{c^2}\delta _{ij},  \label{26}
\end{equation}
or in combination with Eq~(\ref{25}),
\begin{equation}
j_i^s=\rho _{ij}^s[(c^2/\mu )v_j^s-v_j^n]=
\rho _{ij}^s[v_j^{ts}-v_j^n].
\label{27}
\end{equation}
Inserting this expression into Eq~(\ref{max1})
with $\dot{\mathbf{D}}=0$,
taking another curl on both sides, and denoting the
matrix $\rho _{ij}^s$ as
$\boldsymbol{\hat\rho}^s$, we finally obtain
\begin{eqnarray}\nonumber
{\boldsymbol\nabla }\times \left[
\frac{m_{\mathrm{e}}c}e(\boldsymbol{\hat\rho}^s)^{-1}
{\boldsymbol\nabla }\times \mathbf{H}\right] =
{\boldsymbol\nabla }\times (\boldsymbol
v^{ts}-\boldsymbol v^n) \\ =-\frac
e{m_{\mathrm{e}}(1+\zeta)c} {\mathbf
B}-2\mbox{\boldmath$\Omega$}.\label{final}
\end{eqnarray}
This is the equation that accounts for the equilibrium
behavior of the magnetic field in superconductors:
$(m_{\mathrm{e}}c/e)(\boldsymbol{\hat{\rho }}^s)^{-1}$
yields the square of the inverse penetration depth, a
quantity that depends on the crystal direction, while the
vanishing of the right hand expression gives the bulk
value of the magnetic field, Eq~(\ref{0}). Because of
Eq~(\ref{MM}), we may generally neglect the factor
$1+\zeta $.
\section{The Square-Well Potential}
\label{3} Consider a square-well potential, Fig 2, a
popular model for metals. Taking the outside value of the
energy as zero, the depth of the well is $eV(<0)$, with
$V$ the mean-field potential for electrons; filling the
potential up to the Fermi energy $\varepsilon _F(>0)$,
the gap still remaining may be identified as the work
function of the electron $W(<0)$. Together, they satisfy
\begin{equation}
W-\varepsilon _F=eV.  \label{epsi}
\end{equation}
When making contact with thermodynamics, there are no
doubts that the identification
\begin{equation}
W=\tilde{\mu}_{-}  \label{w-mu}
\end{equation}
holds, because the physical significance of $W$ and
$\tilde{\mu}_{-}$, the energy needed to take out an
electron, and the energy gained when putting one in,
respectively, is the same. The other two scales, the
Fermi energy $\varepsilon _F$ and the mean-field
potential $V$, are model-dependent quantities, with a
large measure of arbitrariness attached to them. So we
should not expect them to be directly measurable, or
connected to thermodynamic quantities in a simple
fashion. Nevertheless, all mentioned authors employing
the microscopic approach to understand the London
Moment~\cite{pwa,brady,gutfr,peskin,baym} adopt the
identification
\begin{equation}
V=\Delta \Phi ,  \label{!?}
\end{equation}
with some apparently plausible consequences. First, with
Eqs~(\ref{MM}) and (\ref{19}), we have
$m_{\mathrm{e}}^{*}=m_{\mathrm{e}}(1+\zeta )$ with
\begin{equation} \zeta(m_{\mathrm{e}}c^2)
=\tilde{\mu}_{-}-e\Delta \Phi=W-eV=\varepsilon_F.
\label{mm}
\end{equation}
This makes the mass correction positive,  and rather
tangible: With the estimates $W\approx-4$eV, $eV\approx
-96$eV, and $\varepsilon _F\approx 92$ eV (by averaging
the electron wave functions over the Fermi surface), the
value $\zeta \approx 1.8\times 10^{-4}$ was
found~\cite{gutfr} for the mass correction -- as compared
to $\zeta \approx -10^{-10}$ of the last section.
Moreover, with $\varepsilon _F=\frac
12m_{\mathrm{e}}v_F^2$, Eq~(\ref{mm}) delivers the
simple, kinematic-relativistic interpretation for the
mass correction,
\begin{equation}
\zeta =\textstyle\frac 12(v_F/c)^2\approx 1.8\times
10^{-4}. \label{kr}
\end{equation}
So why do we claim that Eqs~(\ref{!?},\ref{mm},\ref{kr})
are in error? The identification of Eq~(\ref{!?}) is made
by taking the macroscopic electrostatic potential $\Delta
\Phi $ as the potential felt by superconducting electrons
from \textit{``all charge distributions in the metal,
such as surface dipoles..., the screening hole..., and
charge inhomogeneities associated with atomic cores and
valence electrons''}~\cite {brady}. (Sentences in italic
are quotes, here and below.) %Moreover, $\Delta \Phi $ is
%assumed to change rapidly at the surface of the metal,
%within a small depth of a few \AA.
Clearly, the one ostentatiously lacking in this list is
the contribution from the other band electrons (including
the superconducting ones), to the potential, which brings
the potential energy, gained by the last electron to be
added to the system, to
\begin{equation}
W=\tilde{\mu}_{-}=e\Delta \Phi,  \label{last}
\end{equation}
cf Eq~(\ref{7b}). This is the thermodynamic definition of
the potential $\Delta \Phi $, see the discussion of the
last section, and we are not at liberty to alter it. In
fact, as already mentioned, it would have been highly
surprising for the directly measurable macroscopic
electrostatic potential $\Phi $ that forms a 4-vector
with the vector potential $\mathbf{A}$ to be simply
related to the model-dependent mean-field potential $V$.

\begin{figure}
\begin{center}
\includegraphics[width=0.6\columnwidth]{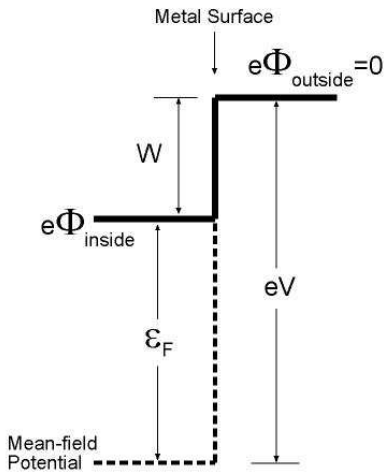}
\end{center}
\caption{The relations between the macroscopic
electrostatic potential $e\Phi$, the work function $W$,
the Fermi energy $\varepsilon_F$, and the mean-field
potential $eV$, close to a metal surface.}\label{Fig2}
\end{figure}

This still leaves the kinetic-relativistic interpretation
of the Fermi velocity, Eq~(\ref{kr}), intact. As it is a
rather popular preconception, and employed by two more
recent papers without the detour via the Fermi
energy~\cite{fischer}, we need to emphasize here that it
is really only correct for free particles, and does not
work for strongly interacting systems.

To understand this, consider an atom, consisting of an
electron and an ion -- the same way the metal consists of
band electrons and the rest. We may write the energy of
the atom as  $E=(m_{\mathrm{i}
}^{*}+m_{\mathrm{e}}^{*})c^2$, or
\begin{equation}
(m_{\mathrm{i}}^{*}+m_{\mathrm{e}}^{*}-m_{\mathrm{i}}-
m_{\mathrm{e}})c^2=W, \label{defect}
\end{equation}
where the superscript $^{*}$ denotes the effective mass
including the mass defect, while $W$ the binding energy
between the electron and the ion, the analogue of the
work function above. Note that the binding energy $W$
includes all contributions, especially the kinetic energy
of the electron. Microscopically, we may divide $W$
between $m_{\mathrm{i}}^{*}$ and $m_{\mathrm{e}}^{*}$
arbitrarily. But if the division is in proportion to the
rest mass, a natural (if not obvious) procedure, the mass
defect of the electron is
\begin{equation}
m_{\mathrm{e}}^{*}=m_{\mathrm{e}}\left( 1+\frac
W{m_{\mathrm{e}}c^2}\frac{m_{
\mathrm{e}}}{m_{\mathrm{i}}+m_{\mathrm{e}}}\right) ,
\label{defect2}
\end{equation}
in agreement with the result of the last section, see
Eqs~(\ref{6a},\ref{7f},\ref{MM}).

Baym considered the problem of mass correction within the
Landau theory of Fermi liquid in his work\cite{baym}. The
theory is beyond the usual mean-field approach, and
accounts for many-body effects. Because his arguments and
calculations are complicated, we shall only quote his
final expression,
\begin{equation}
(m^*_{\mathrm{e}}-m_{\mathrm{e}})c^2=
\varepsilon_{kin}+\varepsilon_{int}-eV_c+\cdots,  \label{mass-baym}
\end{equation}
where $\varepsilon_{kin}$ is the kinetic energy of a
Fermi electron, similar to $\varepsilon_{F}$ above, while
the last two terms are new: $V_c$ is given by
\textit{``the average of the electrostatic potential
within a given unit cell} (eg the Wigner-Seitz cell),
\textit{arising from the charges within the cell."} And
$\varepsilon _{int}$ is \textit{``the interaction energy
of an electron at the Fermi surface, containing
electrostatic as well as exchange contributions} (but
excluding the boundary dipole layer)." The value of
$eV_c$ was given as $-19.8eV$, while the value of
$\varepsilon _{int}$ remains unknown. Because
$\varepsilon _{int}$ not only includes many-body
interactions, but also electrostatic contribution, it may
well be an appreciable term, canceling other
contributions to yield virtually vanishing total
correction, in agreement with the thermodynamic result.
Baym also points out that corrections from the presence
of the lattice are to be expected, because the Landau
Fermi liquid theory is confined to translationally
invariant systems, of which metal electrons are not one.
Only a generalized Landau theory that also considers the
ions would indeed be dealing with a system that is, in
its totality, translationally invariant. This should
result in more generally valid relations, replacing
relations such as  $\boldsymbol{g}= [\mu_-/(m_{\mathrm{e
}}c^2)] \boldsymbol{j}_s$ as given in~\cite{baym}, with
$\boldsymbol{g}= [\mu_{\mathrm{a}}/(m_{\mathrm{e}}
+m_{\mathrm{i}})c^2] \boldsymbol{j}_s$, or Eq~(\ref{23}).

\section{The Boundary Double Layer}
\label{4} Because of the presence of surface dipoles, the
electrostatic potential is discontinuous at the surface
of the superconductor,  $\Delta\Phi\not=0$. If the metal
is in rotation, the vector potential will also be
discontinuous, contributing to the total flux of the
system in the laboratory frame. Since the magnitude of
this is $\sim\Delta\Phi$, taking $e\Delta\Phi$ either as
$\mu_-=W\approx-4$eV, Eq~(\ref{7b},\ref{7a}), or as
$eV\approx-96$eV, Eq~(\ref{!?}), clearly makes a big
difference. As argued at length in the last two sections,
we believe the first is correct, as does the
book~\cite{ll8} by Landau and Lifshitz, who unequivocally
pronounced the equality between the work function and the
potential's discontinuity, $W=e\Delta\Phi$, see \S 23.

The jump of the vector potential follows from the Lorentz
transformation property of the electromagnetic
4-potential $A_\alpha $. Consider a frame co-moving with
the metal. If $A_0^{\prime }$ denotes the electric
potential outside the metal, the potential inside the
metal will be $A_0^{\prime }-\Delta \Phi $. So the
4-potential $A_\alpha ^{^{\prime }}$ in the local rest
frame is $(A_0^{^{\prime }},\boldsymbol{A}^{^{\prime }})$
outside the metal, and $(A_0^{^{\prime }}-\Delta \Phi
,\boldsymbol{A}^{\prime })$ inside it. In the laboratory
frame, the 4-potential $A_\alpha $ becomes, in linear
order of $v/c$,
\begin{equation*}
(A_0^{^{\prime }}-\boldsymbol{v\cdot A}^{^{\prime }}/c,
\boldsymbol{A}^{^{\prime}}-A_0^{^{\prime}}\boldsymbol{v}/c)
\end{equation*}
outside the metal, and
\begin{equation*}
(A_0^{^{\prime }}-\Delta \Phi -\boldsymbol{v\cdot
A}^{^{\prime }}/c,\boldsymbol{A} ^{^{\prime
}}-A_0^{^{\prime }}\boldsymbol{v}/c+\Delta \Phi
\boldsymbol{v}/c)
\end{equation*}
inside the metal, where $\boldsymbol{v}$ is the velocity
of the boundary. So the discontinuity in the vector
potential is
\begin{equation}
\Delta \boldsymbol{A}\equiv
\boldsymbol{A}^{out}-\boldsymbol{A}^{int}=-(\Delta \Phi
/c)\boldsymbol{v}  \label{c3}
\end{equation}
For a metal cylinder uniformly rotating about the z-axis
$\mathbf{\hat{e}_z}$, we have ${\boldsymbol v}= \Omega
\hat{\mathbf e}_z\times {\boldsymbol R}$, with
$\boldsymbol{R}$ the position vector of the boundary. So
the magnetic flux of a non-superconducting, rotating
metal is
\begin{equation}
\oint \Delta \boldsymbol{A}\cdot \mathrm{d} {\mathbf
s}=-2\pi R^2\Delta \Phi \Omega/c.\label{surf-flux}
\end{equation}
Since the experiment~\cite{cabr} measures both this
boundary flux and the London field simultaneously, it is
convenient to introduce an observed mass
$m_{\mathrm{obs}}$ (if the effect due to the penetration
depth is negligible, as is the case in~\cite{cabr}). We
write it as
\begin{equation}
m_{\mathrm{obs}}=(1+\zeta ) (1-\alpha
)m_{\mathrm{e}}=(1-\alpha )m_{\mathrm{e} }^{*}
\label{obs-mass}
\end{equation}
where $\alpha $ accounts for the flux  from the boundary
double layer. In the cited microscopic
considerations,~\cite{pwa,baym,cabr} it is the Fermi
energy, $\varepsilon _F\approx 92$eV, that enters $\zeta
$, and the mean-field potential, $eV\approx -96$eV, that
enters $\alpha $, though in opposite directions.
Combined, the observed mass is corrected only by the work
function $W=eV+$ $\varepsilon _F\approx -4$eV, cf
Eq~(\ref{epsi}), leading to a total correction of about
$8\times 10^{-6}$. Thermodynamically, although $\zeta
\approx -10^{-10}$ is negligible, but the work function
$W$ does enters $\alpha$, so the measured effect is again
given by a value around $-4$eV.

Since the same value for $m_{\mathrm{obs}}$ is predicted
by all theories, one may conjecture that the values of
$\zeta $ or $\alpha $ individually are unimportant,
because any experiment can only observe
$(m_{\mathrm{obs}}/m_{\mathrm{e}})-1$ $=\zeta -\alpha$.
This does not seem right to us. First, both the strength
of surface double layer of a metal and the London field
are well defined physical quantities. And the parameters
$\zeta $ and $\alpha $ are unambiguously related to the
chemical potential and the work function, respectively.
Second, both effects are (at least in principle)
independently measurable. The London field could be
observed directly by measuring the flux near the
top-center of a rotating superconductor (see Fig. 1), and
the flux from the double layer only may be observed in
the normal state, or even in the superconducting state,
by the method given in the next chapter.

\section{A Proposed Experiment}\label{exp}
Consider two concentric, co-rotating, hollow cylinders,
made of the same superconducting metal. The electric
current $J$ flowing through the inner cylinder is
measured by a SQUID, see Fig~3. The cylinders rotate with
the angular velocity $\Omega$. We calculate the magnetic
field and the current in the apparatus. For simplicity we
assume that the cylinders are of infinite height, and no
surface charges (or electric field) are present.

\begin{figure}
\begin{center}
\includegraphics[width=0.6\columnwidth]{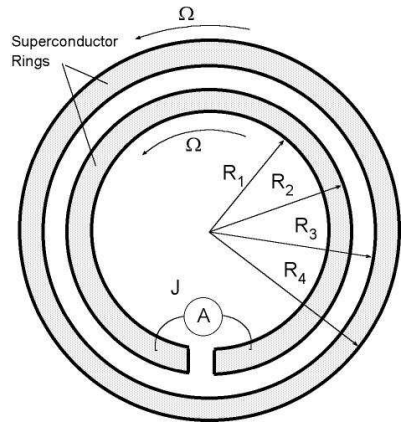}
\end{center}
\caption{Apparatus for measuring the boundary flux in the
superconducting state. Superconducting metal is depicted
by shaded area.}\label{Fig3}
\end{figure}

This geometry was firstly analyzed by Brady~\cite{brady},
who estimated the effect of penetration depths $\lambda$
by simply assuming constant magnetic fields in the
regimes: $0\leq r<R_1+\lambda $ and $R_2-\lambda
<r<R_3+\lambda $, in other words, by assuming that the
actual internal spaces, $0<R_1$ and $R_2<r<R_3$, are
increases by $\lambda$ at the superconducting boundaries.
Moreover, because the magnitudes of the field was
estimated by dividing the magnetic flux by the respective
area, the contributions from the boundary layer have also
been neglected.

The current $J$ in the inner cylinder as determined by
Brady is proportional to the observed mass
$m_{\mathrm{obs}}$ and the difference of the angular
velocities of the two cylinders. This was given as a
method to measure the London field. In our case, both
angular velocities are the same, so $J$ would vanish in
his approximation. Our more detailed calculation,
however, shows that $J$ is finite and observable if the
distance between the two cylinders, $R_3-R_2$, is small
enough. In fact, it is proportional to $\alpha \Omega$,
rendering $\alpha $, or the flux from the boundary layer,
measurable.

The magnetic fields and the current density inside the
cylinders can be obtained by solving Eq~(\ref{final}) in
the cylindrical symmetry. Considering the isotropic case
$\rho _{ij}^s=\varrho _s\delta _{ij}$ with $\varrho _s$
the superfluid density, we write the vector fields as:
\begin{eqnarray}
{\mathbf B} =B\boldsymbol{e}_z,\ {\mathbf
A}=A(\boldsymbol{e}_z\times \boldsymbol{r}/r), \notag \\
\boldsymbol{j}^{\mathrm{e}}
=j^{\mathrm{e}}(\boldsymbol{e}_z\times \boldsymbol{r}/r).
\end{eqnarray}
The general solution  for $\mathbf{H}=\mathbf{B}$ is
\begin{equation}
B=2\gamma ^{*}\Omega +C_1I_0+C_2K_0.  \label{B-sol}
\end{equation}
Notations: $\gamma ^{*}\equiv-m_{\mathrm{e}}^{*}c/e$;
$C_{1-5} $ are integration constants; $\lambda
=(m_{\mathrm{e}}c/\left| e\right| )\sqrt{(1+\zeta
)/\varrho _s}$ is the penetration depth; $I_N,K_N$
($N=0,1$) are Bessel functions with the arguments
$r/\lambda $; $\phi _0=-\hbar c/2e$ is the quantum
fluxoid.

Inserting (\ref{B-sol}) into Eqs ${\mathbf
B}={\boldsymbol\nabla} \times {\mathbf A}$,
$\boldsymbol{j}^{\mathrm{e}}=c{\boldsymbol\nabla} \times
{\mathbf B}$, and noting the quantization property $\oint
{\boldsymbol\nabla} \varphi \cdot \mathrm{d} {\mathbf
s}=2\pi n$ for the phase in Eq (\ref{19a}), we obtain the
magnetic potential and current density:
\begin{eqnarray}
A &=&\gamma ^{*}\Omega r-n\phi _0/r+\lambda
\left( C_1I_1-C_2K_1\right) , \\
j^{\mathrm{e}} &=&-(c/\lambda )(C_1I_1-C_2K_1).
\end{eqnarray}
The integer $n$ denotes quantum state of the cylinder.
The induction $B$ is constant outside the rings, while
the potential is of the form
\begin{equation}
A=Br/2+C/r
\end{equation}
with $C$ again a constant. Using the above solutions, it
can be easily seen that the field of our apparatus is
\begin{equation}
B=\left\{
\begin{array}{l}
B_1;\ \ \ \ \ \ \ \ \ \ \ \ \ \ \ \ \ \ \ \ \ \ \ \ \ \
 \ \ \ \ \ \ \ \ \ \ \ r<R_1,
\\ 2\gamma ^{*}\Omega +C_1I_0+C_2K_0;\ \ \ \ \ R_1<r<R_2,
\\ B_2;\ \ \ \ \ \ \ \ \ \ \ \ \ \ \ \ \ \
 \ \ \ \ \ \ \ \ \ \ \
R_2<r<R_3, \\ 2\gamma ^{*}\Omega +C_3I_0+C_4K_0;\ \ \ \ \
R_3<r<R_4,
\end{array}
\right.  \label{B-field}
\end{equation}
and the potential is
\begin{equation}
A=\left\{
\begin{array}{l}
B_1r/2;\ \ \ \ \ \ \ \ \ \ \ \ \ \ \ \ \ \ \ \ \ \ \
 \ \ \ \ \ \ \ \ \
 r<R_1, \\ \gamma ^{*}\Omega r-n_1\phi _0/r+\lambda
(C_1I_1-C_2K_1);\\\qquad\qquad\qquad
 \ \ \ \ \ \ \ \ \ \ \  \ \ \ \ \ R_1<r<R_2,
\\
B_2r/2+C_5/r; \ \  \ \ \ \ \ \ \ \ \ \ \ \ \ R_2<r<R_3,
\\ \ \ \ \gamma ^{*}\Omega r-n_2\phi _0/r+\lambda
(C_3I_1-C_4K_1);\\\qquad\qquad\qquad  \ \ \ \ \ \ \ \ \ \
 \ \ \ \ \ \ \ R_3<r<R_4.
\end{array}
\right.  \label{A-field}
\end{equation}
Note that $n_1,n_2$ may be different, because the two
cylinders can be in different quantum states. The current
per unit length in the inner cylinder is
\begin{eqnarray}
J &=&\int_{R_1}^{R_2}j^{\mathrm{e}}\mathrm{d}r  \notag \\
&=&-c[C_1(I_{02}-I_{01})+C_2(K_{02}-K_{01})].  \label{ammeter}
\end{eqnarray}
Here and below we will use the notations
\begin{equation*}
I_{NM}=I_N(R_M/\lambda ),\ \ K_{NM}=K_N(R_M/\lambda ),
\end{equation*}
where the integers $N=0,1$, $M=1-4$. The unknown
constants $C_{1-5}$,  $B_1,B_2$ are determined from the
boundary conditions. At $R_1$, according to
Eqs~(\ref{B-field},\ref{A-field}), we have
\begin{eqnarray}
B_1 =2\gamma ^{*}\Omega +C_1I_{01}+C_2K_{01},
\label{R1-1} \\ B_1R_1/2 =\left( \gamma ^{*}-\Delta \Phi
/c\right) \Omega R_1  \notag \\ -n_1\phi _0/R_1+\lambda
(C_1I_{11}-C_2K_{11}),  \label{R1-2}
\end{eqnarray}
where Eq~(\ref{R1-1})  shows that the induction varies
continuously at the surface. Eq~(\ref{R1-2}) is obtained
from the jump of the potential Eq~(\ref{c3}). Similarly,
we have at $R_2$:
\begin{eqnarray}
B_2 =2\gamma ^{*}\Omega +C_1I_{02}+C_2K_{02},
\label{R2-1} \\ B_2R_2/2+C_5/R_2 =\left( \gamma
^{*}-\Delta \Phi /c\right) \Omega R_2 \notag \\ -n_1\phi
_0/R_2+\lambda (C_1I_{12}-C_2K_{12}), \label{R2-2}
\end{eqnarray}
and at $R_3$:
\begin{eqnarray}
B_2 =2\gamma ^{*}\Omega +C_3I_{03}+C_4K_{03},
\label{R3-1} \\ B_2R_3/2+C_5/R_3 =\left( \gamma
^{*}-\Delta \Phi /c\right) \Omega R_3 \notag \\ -n_2\phi
_0/R_3+\lambda (C_3I_{13}-C_4K_{13}). \label{R3-2}
\end{eqnarray}
The  induction is zero beyond $R_4$. (We consider the
case of no externally applied field, ie all the sources
of the field are from supercurrents in the rings.) We
have then
\begin{equation}
2\gamma ^{*}\Omega +C_3I_{04}+C_4K_{04}=0.  \label{R4}
\end{equation}
Solving Eq~(\ref{R1-1}-\ref{R4}) for $C_{1-5}$, $B_1,B_2$
and inserting  them into Eq~(\ref{ammeter}), we obtain
the current. The result can be written as
\begin{equation}
J=[f_3n_1+f_4(n_2-n_1)]J_0+c\gamma ^{*} (f_2-f_1\alpha)
\Omega,   \label{J}
\end{equation}
with
\begin{equation}
\alpha =-\frac{e\Delta \Phi }{m_{\mathrm{e}}^{*}c^2},\ \
J_0=-\frac{\hbar c^2 }{2eR_1^2}.  \label{alpha}
\end{equation}
Here the factors $f_{1-4}$ depend only on the penetration
depth and geometry of the system. For observing the
boundary effect, it is convenient to choose the geometry
such that $\left| f_2\right| <<\left| \alpha f_1\right|
$. One example satisfying the requirement is: $R_1=2.5$
\textrm{cm}, $R_2-R_1=1$ $\mathrm{\mu m}$, $R_3-R_2=40$
$\mathrm{nm}$, $R_4-R_3=1$ $\mathrm{\mu m}$, $\lambda
=40$ $\mathrm{nm}$. We have obtained, by the numerical
computation, the values
\begin{eqnarray*}
f_1 &=&1.33333,\ \ f_2=1.852\times 10^{-11}, \\
f_3 &=&-1.9999968,\ \ f_4=2.083\times 10^7.
\end{eqnarray*}
Because $f_4$ is very large, it is not probable for the
two rings to lay in different quantum states. So within a
good approximation, (and also neglecting $\zeta \ll 1$)
we have
\begin{equation}
J=-2J_0n_1-f_1(\Delta \Phi )\Omega,
\end{equation}
with $f_1=4/3$ for the geometry considered. This shows
that the strength of the surface double layer $\Delta
\Phi $ can be measured by measuring the change of the
current with the rotation. Note that the factor $f_1$
decreases with increasing distance $R_3-R_2$ between the
two cylinders, see Fig 4. When the distance is large,
$f_1\rightarrow 0$, the last term in (\ref{J}) is
negligible, and we return to the result by
Brady~\cite{brady}.

\begin{figure}
\begin{center}
\includegraphics[width=0.6\columnwidth]{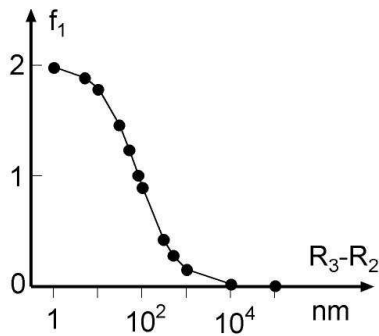}
\end{center}
\caption{Variation of the factor $f_1$ with the
separation distance of the two rings.}\label{Fig4}
\end{figure}

\section{Conclusions}
\label{con} The conclusion of this paper is that  the
relativistic correction to the London field is not the
result of the Fermi velocity. Instead, it is given by the
chemical potential $\mu$ of the metal, which quantifies
the complicated interaction among all the particles,
including that between electrons and ions. Because the
interaction energy is typically several eV per atom,
while the mass of an atom is around $10^4$MeV, the
relativistic correction is tiny, of order $10^{-10}$, and
beyond the scope of any present experimental techniques.
When previous, microscopic theories considered the London
Moment, they considered the kinetic contribution to the
correction $\zeta$ of the electron's mass, but neglected
other contributions, especially from the interaction
between the electrons and the lattice. In addition, the
discontinuity of the macroscopic electrostatic potential
at the metal surface was incorrectly taken to be the mean
field potential, or the sum of the kinetic energy and the
work function. As soon as these errors are revised,
$\zeta$ is found to essentially vanish. Unfortunately,
the reason for the discrepancy between the
experiment~\cite{cabr} and all theories remains unclear.

\end{document}